\documentclass{article}

\usepackage{hyperref}
\usepackage{amssymb}
\usepackage{amsmath}
\usepackage{mathtools}
\usepackage{theorem}
\usepackage{tikz}
\usepackage{microtype}

\usetikzlibrary{calc}

\newtheorem{definition}{Definition}
\newtheorem{theorem}{Theorem}

\newcommand{\mb}[1]{{\bf #1}}
\newcommand{\mc}[1]{{\mathcal{#1}}}

\newcommand{\fall}[1]{{\forall\,{#1},\ }}

\DeclareMathOperator\argtanh{argtanh}

\DeclareMathOperator\dev{dev}

\renewcommand{\d}{\mathrm{d}}

\makeatletter
\newcommand{\raisemath}[1]{\mathpalette{\raisem@th{#1}}}
\newcommand{\raisem@th}[3]{\raisebox{#1}{$#2#3$}}
\makeatother

\begin{document}

\title{Geometric Properties of Paths \\ in Relativistic Lagrangian Mechanics}
\author{Olivier Brunet\footnote{\texttt{olivier.brunet} at \texttt{normalesup.org}}}
\maketitle

\abstract Considering an extension of the principle of covarience to the action along a path in relativistic Lagrangian mechanics, we motivate the use of \emph{geometric} -- i.e.\ covariant and parameter invariant -- Lagrangian functions. We then study some properties of geometric Lagrangians, and introduce the notion of \emph{deviation} of a path, which is a covariant measure of how much a path departs from a geodesic. Finally, we apply this notion of the twin paradox, and provide a rigorous resolution of it.

\section{Introduction}

More than a century after the creation of Einstein's special theory of relativity, a clear and definitive formulation of relativistic Lagrangian mechanics remains to be defined: if a relativistic version of Lagrangian mechanics can indeed be found in most textbooks (a far-from-exhaustive list being for instance \cite{Goldstein:ClassicalMechanics,JoseSaletan:ClassicalMechanics,Rindler06:Relativity}), there is no consensus regarding which Lagrangian to use, even for the simple case of a free particule.
%
%
In particular, the question remains whether one should stick to velocity-homogeneous Lagrangian functions (at the loss of Hamiltonian mechanics) or not.

In this article, we introduce the notion of \emph{geometric} Lagrangian functions, by which we mean that it is both covariant and parametrization invariant. We justify the use of this type of Lagrangian function by the necessity for the action along a path to only depend on the geometry of the path, i.e.\ on the path seen as a connected set of timelike-separated spacetime events. As a consequence, the Lagrangian function can neither depend on any reference frame nor on any parametrization. 

We then study some of the properties of geometric Lagrangian. In particular, we show how the application of Euler's homogeneous function theorem -- a consequence of which being that the associated Hamiltonian is trivially null -- leads to the definition of the \emph{deviation} of a path, which is a purely geometric measure of the way a path deviates from a geodesic.

Finally, we show that the deviation of a path has an extremely simple geometric interpretation for a free particle in special relativity, and we use this notion to provide a formal and rigourous resolution of the twin paradox.

\section{Geometric Lagrangians}

The \emph{principle of covariance}, which states that the formulation of physical laws should only rely on those physical quantities which value can be determined unambiguously regardless of the frame of reference, plays a central role in the theory of relativity. It emphasizes the idea that the relevant physical quantities should be purely geometrical instead of analytical.

Let us extend this idea to pathes and, more specifically, let us focus on the Lagrangian-based action along a path~:
\begin{equation} \label{eq:action1}
S(\mc P) = \int_a^b L\bigl(\mb x(t), \dot {\mb x}(t)\bigr) \, \d t 
\end{equation}
Here, we consider that path $\mc P$ is parametrized as $\{ \mb x(t) \mid t \in [a, b] \}$. Consider now another parametrization of the same path, of the form $\{ \mb y(u) \mid u \in [\alpha, \beta] \}$ where $\mb y(u) = \mb x \circ \varphi(u)$, $a = \varphi(\alpha)$, $b = \varphi(\beta)$, and
$$ \fall {u \in [\alpha, \beta]} \varphi'(u) > 0 $$
The action of the path, computes using the second parametrization,~is:
\begin{equation} \label{eq:action2}
S'(\mc P) = \int_\alpha^\beta L\bigl(\mb y(u), \dot {\mb y}(u)\bigr) \, \d u = \int_\alpha^\beta L\bigl(\mb x \circ \varphi(u), \varphi'(u) \, \dot {\mb x} \circ \varphi(u)\bigr) \, \d u
\end{equation}
But considering the change of variable $t = \varphi(u)$ in eq.~\eqref{eq:action1}, we also have:
\begin{equation} \label{eq:action3}
 S(\mc P) = \int_a^b L\bigl(\mb x(t), \dot {\mb x}(t)\bigr) \, \d t = \int_\alpha^\beta L\bigl(\mb x \circ \varphi(u), \dot {\mb x} \circ \varphi(u)\bigr) \, \varphi'(u) \, \d u
 \end{equation}

Extending the principle of covariance to the value of the action, if we want it to depend only one the path but not on its parametrization (i.e.\ it depends only on the set $\{\mb x(t) \mid t \in [a, b]\} = \{\mb y(u) \mid u \in [\alpha, \beta]\}$), we shall have
$$ S(\mc P) = S'(\mc P).$$
In order to have this equality for any path and any reparametrization function $\varphi$ such that~$\varphi' > 0$, this implies that the Lagrangian function has to verify:
$$ \fall {\mb x, \mb v} \fall {\lambda > 0} L(\mb x, \lambda \mb v) = \lambda L(\mb x, \mb v), $$
In other words, $L$ has to be $1$-homogeneous in its velocity argument. We thus define:
\begin{definition}[Geometric Lagrangian]
A Lagrangian function $L$ is \emph{geometric} if it is both covariant and $1$-homogeneous in its second argument.
\end{definition}

As it is well known, having an homogeneous Lagrangian in its velocity argument has dramatic consequences regarding hamiltonian mechanics. If we define the conjugate momentum
$$ p^\mu = \frac {\partial L}{\partial \dot x_\mu} $$
then Euler's homogeneous function theorem entails
$$ p_\mu x^\mu = L $$
so that the associated Hamiltonian obtained as the Legendre transform of $L$~is
$$ H = p_\mu x^\mu - L = 0. $$
Moreover, since Euler's theorem states an equivalence, this means that a Lagrangian function is precisely parameter independent if and only if the associated Hamiltonian is constantly zero.

\ 

However, we insist on the idea that it is extremely important to consider geometric Lagrangian, as it is a continuation of Einstein's effort to to formulate physical theories in a covariant way, reflecting the idea that coordinates and related notions, such as reference frames, are only artifacts used for describing nature in an analytical way. To that respect, the previous remark shows that hamiltonian mechanics is not compatible with the requirement of covariance.

This position is obviously not new, and was also advocated for instance by Dirac \cite{Dirac33:Lagrangian}: ``there are reasons for believing that the Lagrangian [formulation] is the more fundamental [than the Hamitonian one. In particular,] the Lagrangian method can easily be expressed relativistically, on account of the action function being a relativistic invariant; while the Hamiltonian method is essentially non-relativistic in form, since it marks out a particular time variable as the canonical conjugate of the Hamiltonian function.''

\ 

In the next sections, we will explore some of the benefits of using geometric Lagrangian functions.

\section{The Design of Lagrangian Functions}

Let us first explore how the necessity of having a geometric Lagrangian provides guidelines for designing Lagrangian functions. It can first be remarked that any linear combination of terms of the form
$$ \sqrt{\dot x_\mu \dot x^\mu}, \ A_\mu(\mb x) \dot x^\mu, \ \frac 1 {\sqrt{\dot x^\mu \dot x_\mu}} B_{\nu \eta}(\mb x) \dot x^\nu \dot x^\eta, \ \frac 1 {\dot x^\mu \dot x_\mu} C_{\nu \eta \kappa}(\mb x) \dot x^\nu \dot x^\eta \dot x^\kappa, \ \hbox{etc.} $$
leads to a geometric Lagrangian function. 

Consider, for instance, the case of a free particle. The literature provides a large variety of relativistic Lagrangians. For instance, taken from \cite{Goldstein:ClassicalMechanics,JoseSaletan:ClassicalMechanics,Rindler06:Relativity,HobsonEfstathiouLasenby:GR}, up to a multiplicative scalar constant, it is possible to find:
$$
\dot x_\mu \dot x^\mu \qquad \qquad \sqrt{\dot x_\mu \dot x^\mu} \qquad \qquad \sqrt{1 - \frac {\dot x_i \dot x^i}{c^2}} = \sqrt{1 - \beta^2} $$
where $(\dot x_i)$, with $i$ ranging from $1$ to $3$, represents the 3-speed of a particle in the Lorentz frame ``under consideration''. 
The presence of an additional (time independent) potential energy usually leads to an additional term of the form $ - U(x, y, z) $, providing to a Lagrangian such~as
\begin{equation} \label{eq:lagaga}
- m  c^2 \sqrt{1 - \beta^2} - U(x,y,z)
\end{equation}

However, quoting \cite{JoseSaletan:ClassicalMechanics}, ``this treatment of the relativistic particle exhibits some of the imperfections of the relativistic Lagrangian (and Hamiltonian) formulation of classical dynamical systems. For one thing, it uses the nonrelativistic three-vector velocity and position but uses the relativistic momentum. For another, all of the equations are written in the special coordinate system in which the potential is time independent, and this violates the relativistic principle according to which space and time are to be treated on an equal footing.'' We do agree with these objections, and the requirement of having a geometric Lagrangian can help to select the correct form.

Considering the kinetic term alone, it is clear that $\sqrt{1 - \beta^2}$ is not covariant, and that $\dot x_\mu \dot x^\mu$ is not 1-homogeneous so that the only reasonnable candidate for a geometric Lagrangian is (with the correct multiplicative constants)
$$ - m c \sqrt{\dot x^\mu \dot x_\mu} $$
even though it might look ``awkward'' \cite{Rindler06:Relativity}.

Now, regarding potential energy, a term of the form $-U(x, y, z)$ is clearly not suitable for a geometric Lagrangian. It is, in particular, not 1-homogeneous in $\dot {\mb x}$. But it can easily be turned into a suitable geometric form the following way. Let~$\mb e_\mu$ be a vector basis for the Lorentz frame in which $U$ is time-independent. As $U$ represents an energy, this suggests to see it as the time-component of a 4-vector and, indeed, if one defines:
$$ \mb A(c \, t, x, y, z) = \frac {U(x, y, z)} {\raisemath{1pt}c} \mb e_0, $$
it is then easy to verify that the term $ - A_\mu(\mb x) \, \dot x^\mu $ leads to the correct equation of motion using the geometric Lagrangian
\begin{equation} \label{eq:pot_geo}
- m c \sqrt{\dot x_\mu \dot x^\mu} - A_\mu(\mb x) \, \dot x^\mu = 
- m c \sqrt{\dot {\mb x} \cdot \dot {\mb x}} - \frac {U(\mb x)} {\raisemath{1pt}c} \, \mb e_0 \cdot \dot {\mb x}
\end{equation}
For instance, in $1+1$ dimension, for a scalar potential $V(x)$, the Lagrangian becomes:
$$ L(t, x, t', x') = - m c \sqrt{c^2 t'^2 - x'^2} - V(x) \, t'$$
where primed quantities correspond to their derivative w.r.t.\ the path parameter. It can be remarked that if factoring by $t'$, one can recognize the Lagrangian of equation~\eqref{eq:lagaga}:
$$ L(t, x, t', x') = t' \times \Biggl(- m c^2 \sqrt{1 - \frac 1 {c^2} \Bigl(\frac {x'}{t'}\Bigr)^2} - V(x) \Biggr) $$

\section{The Euler-Lagrange Vector}

Let us return now to the equality verified by a geometric Lagrangian as follows from Euler's homogeneous function theorem:
$$ L = \frac {\partial L}{\partial \dot {\mb x}} \cdot \dot {\mb x} $$
If we differentiate it w.r.t.\ the path parameter~$t$, we obtain
\begin{equation} \label{eq:elvector}
\Bigl(\frac {\partial L}{\partial \mb x} - \frac {\d}{\d t} \frac {\partial L}{\partial \dot {\mb x}}\Bigr) \cdot \dot {\mb x} = 0
\end{equation}
In the first factor of the product, we recognize the term which, in the Euler-Lagrange equation, is supposed to equal $0$:
\begin{equation}
\frac {\partial L}{\partial \mb x} - \frac {\d}{\d t} \frac {\partial L}{\partial \dot {\mb x}} = \mb 0 \label{(EL)}
\end{equation}
However, as $L$ is covariant, this quantity is actually a $4$-vector, and equation~\eqref{eq:elvector} tells us that it is a spacelike one, as $\dot {\mb x}$ is timelike and $\dot {\mb x} \neq 0$. In particular, we have
$$ \Bigl\| \frac {\partial L}{\partial \mb x} - \frac {\d}{\d t} \frac {\partial L}{\partial \dot {\mb x}} \Bigr\| = 0 \iff \frac {\partial L}{\partial \mb x} - \frac {\d}{\d t} \frac {\partial L}{\partial \dot {\mb x}} = \mb 0 $$
This suggest the following definition:
\begin{definition}[Deviation of a Path]
Given a geometric Lagrangian $L$ and a smooth path $\mc P$, we define its \emph{deviation}~as
$$ \dev(\mc P) = \int_{\mc P} \Bigl\| \frac {\partial L}{\partial \mb x} - \frac {\d}{\d t} \frac {\partial L}{\partial \dot {\mb x}} \Bigr\| \d t $$
\end{definition}
Then, as a direct consequence of the Euler-Lagrange equation, we have:
\begin{theorem}
A path $\mc P$ is a geodesic w.r.t.\ a geometric Lagrangian $L$ iff $\dev(\mc P) = 0$.
\end{theorem}
Moreover, it is easy to prove~that

\begin{theorem}
The deviation of a path is a geometric quantity.
\end{theorem}

\medskip

It can be remarked finally that the Euler-Lagrange vector attached to a point of a path is not geometric, as its norm depends on the parametrization. However, dividing it by $\| \dot {\mb x}\|$ leads to a geometric vector attached to each point of a path. Formally, we define:

\begin{definition}[Euler-Lagrange Vector]
The Euler-Lagrange attached to a point of a path is
$$ \frac 1 {\sqrt{\dot{\mb x} \cdot \dot{\mb x}}} \Bigl(\frac {\partial L}{\partial \mb x} - \frac {\d}{\d t} \frac {\partial L}{\partial \dot {\mb x}}\Bigr) $$
or, more precisely, if a path $\mc P$ is parametrized as $\bigl\{\mb x(t) \bigm| t \in [a, b]\bigr\}$, then the Euler-Lagrange vector at $\mb x(t)$~is\footnote{where $\partial_1 L$ (resp. $\partial_2 L$) denotes $\frac {\partial L}{\partial \mb x}$ (resp. $\frac {\partial L}{\partial \dot {\mb x}}$)}
$$ \frac 1 {\| \dot {\mb x}(t)\|} \Bigl(\partial_1 L\bigl(\mb x(t), \dot {\mb x}(t)\bigr) - \bigl[u \mapsto \partial_2 L\bigl(\mb x(u), \dot {\mb x}(u)\bigr)\bigr]'(t) \Bigr)$$
\end{definition}
It is easy to check 
that:
\begin{theorem}
The Euler-Lagrange vector attached to a point of a path is a geometric quantity.
\end{theorem}
The previous definition of the deviation of a path $\mc P$ can then be rewritten, in terms of the Euler-Lagrange vector,~as
$$ \dev(\mc P) = \int_{\mc P} \frac 1 {\bigl\| \dot {\mb x}\bigr\|} \Bigl\| \frac {\partial L}{\partial \mb x} - \frac {\d}{\d t} \frac {\partial L}{\partial \dot {\mb x}} \Bigr\| \d s $$
with $\d s(t) = \| \dot {\mb x}(t)\| \, \d t$. 

\section{The Geometric Lagragian of a Free Particle}

Having defined the deviation of a path, let us now explore this notion in the case of the geometric lagrangian of a free particle:
$$ L(\mb x, \dot {\mb x}) = - m c \bigl\| \dot {\mb x}\bigr\| = - m c \sqrt{\dot {\mb x} \cdot \dot {\mb x}} = - m c \sqrt{\dot x_\mu \dot x^\mu} $$
We then have $\partial_1 L(\mb x, \dot {\mb x}) = \mb 0$,~and
$$ \partial_2 L(\mb x, \dot {\mb x}) = - \frac {m c}{\| \dot {\mb x}\|} \dot {\mb x} $$
%
%
so that the corresponding Euler-Lagrange vector~is
$$ \frac 1 {\sqrt{\dot {\mb x} \cdot \dot {\mb x}}} \Bigl(\partial_1 L - \frac {\d}{\d t} \partial_2 L\Bigr) = - \frac {m c}{\dot {\mb x} \cdot \dot {\mb x}}\Bigl(\ddot {\mb x} - \frac {\dot {\mb x} \cdot \ddot {\mb x}} {\dot {\mb x} \cdot \dot {\mb x}} \dot {\mb x}\Bigr) = - \frac {m c}{\bigl(\dot {\mb x} \cdot \dot {\mb x}\bigr)^2}\Bigl(\bigl(\dot {\mb x} \cdot \dot {\mb x}\bigr) \ddot {\mb x} - \bigl(\dot {\mb x} \cdot \ddot {\mb x}\bigr) \dot {\mb x}\Bigr) $$
and the infinitesimal deviation (the norm of the Euler-Lagrange)~is
$$ \Bigl\| \frac 1 {\sqrt{\dot {\mb x} \cdot \dot {\mb x}}} \Bigl(\partial_1 L - \frac {\d}{\d t} \partial_2 L\Bigr) \Bigr\| = m c \sqrt {\frac{\bigl(\dot {\mb x} \cdot \dot {\mb x}\bigr) \bigl(\ddot {\mb x} \cdot \ddot {\mb x}\bigr) - \bigl(\dot {\mb x} \cdot \ddot {\mb x}\bigr)^2} {\bigl(\dot {\mb x} \cdot \dot {\mb x}\bigr)^3}} $$
One recognizes, up to the ``$mc$''-factor, the curvature $\gamma$ of the path. In other worlds, the deviation of a path for a free particle is, up to the previous factor, the mere integral along the path of its curvature:
$$ \dev(\mc P) = m c \! \! \int_a^b \! \! \gamma(t) \| \dot {\mb x}(t) \| \, \d t $$

\section{Change of velocity, and the Twin Paradox}

The previous remark leads to the following point: suppose again that a path $\mc P$ is parametrized as~$\bigl\{\mb x(t)\bigm| t \in [a, b]\bigr\}$. Its velocity evolves from $\frac {\dot {\mb x}(a)}{\|\dot {\mb x}(a)\|}$ to $\frac {\dot {\mb x}(b)}{\|\dot {\mb x}(b)\|}$, so that its deviation is \emph{at least}
$$ m c \mathop{\mathrm{argcosh}}\Bigl(\frac {\dot {\mb x}(a)}{\|\dot {\mb x}(a)\|} \cdot \frac {\dot {\mb x}(b)}{\|\dot {\mb x}(b)\|}\Bigr). $$
For instance, if in some orthonormal basis $(\mb e_\mu)$, one has $\dot {\mb x}(a) = \mb e_0$, and $\dot {\mb x}(b) = \mb e_0 + \dfrac v {\raisemath{1pt}c} \, \mb e_1$, this quantity amounts~to
$$ 
m c \mathop{\mathrm{argcosh}}\Biggl(\frac 1 {\sqrt {1 - \frac {v^2}{c^2}}}\Biggr) = m c \mathop{\mathrm{argtanh}} \frac v {\raisemath{1pt}c} $$
and this minimum can easily be obtained, for instance with
$$ \mb x(t) = \sin t \, \mb e_0 + \cos t \, \mb e_1 \qquad \qquad \hbox{where $t \in \bigl[0, \arctan \frac v {\raisemath{1pt}c}\bigr]$}, $$
or with
$$ \mb x(t) = c \sinh(t) \, \mb e_0 + v \cosh t \, \mb e_1  \qquad \qquad \hbox{where $t \in [0, +\infty]$}. $$
It can be noted, though, that the deviation does not depend on the particulars of the path, as long as it is ``reasonable'', i.e.\ if, up to a norm factor, $\dot {\mb x}(t)$ can be written as $\mb e_0 + f(t) \mb e_1$, with $f$ increasing from $0$ to $\dfrac v {\raisemath{1pt}c}$. To sum this up,

\begin{theorem}
A change of velocity of $\Delta v$ induces a deviation increase of $m c \argtanh \Bigl( \dfrac {\Delta v} {\raisemath{1pt}c} \Bigr)$.
\end{theorem}

This result leads to a simple formal resolution of the famous ``Twin Paradox'' \cite{langevin11:twins}. We recall that this paradox involves two twins, one of whom makes a journey into space in a high-speed rocket and returns home to find that his twin, who has remained on Earth, has aged more. The puzzling aspect of this result is that there seems to be a symmetry between the two twins, each of them seeing the other as moving w.r.t.\ himself. However, this interpretation is rather  naive, as one twin remains in a single inertial frame while the trajectory of the other twin, the one in the rocket, involves at least two reference frames: one for the outbound journey and another for the inbound~one. But expressed this way, this distinction remains rather unclear and still demands a more rigourous formalization.

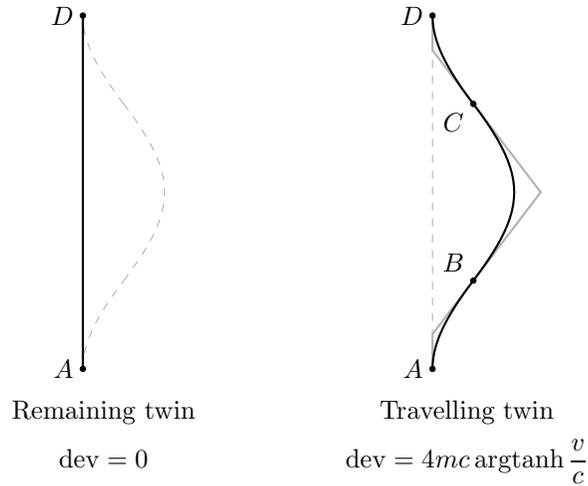
\begin{figure}
\begin{centering}
\begin{tabular}{c@{\hspace{2cm}}c}
\begin{tikzpicture}
\coordinate (A) at (0, 0);
\coordinate (D) at (0,4.700804775) ;

\draw[domain=0:1, dashed, black!30] plot ({cosh(\x)-1},{sinh(\x)});
\draw[domain=-1:1, dashed, black!30] plot ({2.086161270-cosh(\x)},{sinh(\x)+2.350402387});
\draw[domain=-1:0, dashed, black!30] plot ({cosh(\x)-1},{4.700804775+sinh(\x)});


\draw [fill] (A) circle (1pt) ;
\draw [fill] (D) circle (1pt) ;

\node [left] at (A) {$A$} ;
\node [left] at (D) {$D$} ;

\draw [thick] (A) -- (D) ;
\end{tikzpicture} &
\begin{tikzpicture}
\coordinate (A) at (0, 0);
\coordinate (AB) at (0, 0.4621171573) ;
\coordinate (B) at (0.5430806348, 1.175201194);
\coordinate (BC) at (1.438106996, 2.350402387) ;
\coordinate (C) at (0.5430806348, 3.525603581) ;
\coordinate (CD) at (0, 4.2386876173) ;
\coordinate (D) at (0,4.700804775) ;

\draw [dashed, black!30] (A) -- (D) ;

\draw [black!30,thick] (A) -- (AB) -- (BC) -- (CD) -- (D) ;

\draw [fill] (A) circle (1pt) ;
\draw [fill] (B) circle (1pt) ;
\draw [fill] (C) circle (1pt) ;
\draw [fill] (D) circle (1pt) ;

\node [left] at (A) {$A$} ;
\node [above left] at (B) {$B$} ;
\node [below left] at (C) {$C$} ;
\node [left] at (D) {$D$} ;

\draw[domain=0:1, thick] plot ({cosh(\x)-1},{sinh(\x)});
\draw[domain=-1:1, thick] plot ({2.086161270-cosh(\x)},{sinh(\x)+2.350402387});
\draw[domain=-1:0, thick] plot ({cosh(\x)-1},{4.700804775+sinh(\x)});
\end{tikzpicture} \\
Remaining twin & Travelling twin \\[4pt]
$\dev = 0$ & $\dev = 4 m c \argtanh \dfrac v {\raisemath{1pt}c}$
\end{tabular}

\end{centering}
\caption{Twin Paradox} \label{fig:twin3}
\end{figure}

The notion of path deviation actually provides a very simple and general way to quantify this idea of ``involving two frames'' and thus to distinguish between both routes, which are depicted on figure~\ref{fig:twin3}. Regarding the twin who remains on Earth, if we neglect the movements she makes on Earth, the deviation of its path is~$0$. The situation of the second twin is rather different:
\begin{enumerate}
	\item Between $A$ and $B$, there is an acceleration phase, resulting of a final velocity of $\dfrac v {\raisemath{1pt}c}$ outward;
	\item between $B$ and $C$, there is an earthward acceleration, resulting of a reversal of its velocity;
	\item finally, between $C$ and $D$, there is a deceleration phase resulting to an arrival on Earth with velocity zero.
\end{enumerate}
These three phases contribute to the overall deviation of, respectively,
$$ m c \argtanh \dfrac v {\raisemath{1pt}c}, \qquad 2 m c \argtanh \dfrac v {\raisemath{1pt}c} \qquad \hbox{and} \qquad m c \argtanh \dfrac v {\raisemath{1pt}c}$$
so that this path has an overall deviation of $4 m c \argtanh \dfrac v {\raisemath{1pt}c}$ (and still half as much if one only takes the second phase into account).

Again, the deviation of the path does not depend of its exact shape, as long as the journey can be decomposed in the three previous phases: acceleration outward up to velocity $\frac v {\raisemath{1pt}c}$, change of direction so as to head towards Earth at velocity $\frac v {\raisemath{1pt}c}$ and finally braking phase. Moreover, as deviation of a path is a purely geometric quantity, this deviation does obviously not depend on any reference frame or parametrization of the path.

\section{Concluding Remarks}

The notion of \emph{geometric} Lagrangian, if not new, is still to be widely accepted, as illustrated by the previous textbook examples. Yet, the principe of covariance applied to paths implies that the properties of covariance and parametrization invariance are a necessity for a Lagrangian function if one considers that a path is anything but a connected collection of spacetime events. Of the two properties, the requirement of covariance is widely accepted. Parametrization invariance, on the other hand, appears to be more problematic. The reason is that this is equivalent for the Lagrangian function to be homogenous in its velocity argument. But then, because of Euler homogeneity theorem, this implies that the associated Hamiltonian is trivially null. As a consequence, using geometric Lagrangians implies that one must entirely renounce the use of Hamiltonian mechanics, which might be seen as too much a price to pay.

But again, the principle of covariance -- which leads to parametrization invariance of Lagrangian functions -- constitutes, in our opinion, a strong enough requirement for preferring it over the possibility of Hamiltonian mechanics.

The notion of path deviation illustrates the advantage of considering geometric Lagrangians. It leads to a direct characterization of the solutions of the Euler-Lagrange equation, by demanding that their deviation be zero. More generally, it provides a measure of how much a path departs from a geodesic (i.e.\ a solution of the Euler-Lagrange equation) and the example of the twin paradox shows the usefulness of such a measure.

\ 

Finally, we would like to mention a domain in which the deviation of a path might prove useful, namely Feynman's path integrals~\cite{Feynman48,FeynmanHibbs:PathIntegral,KleinertPathes}. We recall that they constitute a Lagrangian-based formulation of quantum mechanics: In a non-relativistic setting, the probability amplitude for a particule to go from a point to another is given by adding the contribution of \emph{all} paths between these two points. Quoting \cite{Feynman48},
\begin{quotation}
[Postulate II:] The paths contribute equally in magnitude, but the phase of their contribution is the classical action (in units of $\hbar$); i.e., the time integral of the Lagrangian taken along the path.
\end{quotation}
In a relativistic setting, the situation is more complicated, and it is usually agreed that one cannot preserve both covariance and causality (i.e.~only considering pathes that go forward in time in any Lorentz frame)~\cite{Redmount-RQMPath,teitelboim1983}. Should then one favor covariance~\cite{henneauxteitelboim} or causality~\cite{HartleKuchar86}? In order to attempt obtaining a formulation of path integrals verifying both covariance and causality, we suggest to amend slightly the previous postulate by assigning possibly different magnitudes to different paths, and we think that a reasonable candidate for this would be to have the magnitude depend on the deviation of the path.

\bibliographystyle{alpha}

\begin{thebibliography}{HEL06}

\bibitem[Dir33]{Dirac33:Lagrangian}
P.~A.~M. Dirac.
\newblock The {L}agrangian in {Q}uantum {M}echanics.
\newblock {\em Physikalische {Z}eitschrift der {S}owjetunion}, 3, 1933.

\bibitem[Fey48]{Feynman48}
Richard~P. Feynman.
\newblock Space-{T}ime {A}pproach to {N}on-{R}elativistic {Q}uantum
  {M}echanics.
\newblock {\em Review of Modern Physics}, 20(367), 1948.

\bibitem[FH65]{FeynmanHibbs:PathIntegral}
Richard~P. Feynman and Albert~R. Hibbs.
\newblock {\em Quantum {M}echanics and {P}ath {I}ntegrals}.
\newblock Dover Edition, 1965.

\bibitem[GPS00]{Goldstein:ClassicalMechanics}
Herbert Goldstein, Charles Poole, and John Safko.
\newblock {\em Classical {M}echanics}.
\newblock Addison-Wesley, 2000.

\bibitem[HEL06]{HobsonEfstathiouLasenby:GR}
Michael~P. Hobson, George~P. Efstathiou, and Anthony~N. Lasenby.
\newblock {\em General {R}elativity, An {I}ntroduction for {P}hysicists}.
\newblock Cambridge University Press, 2006.

\bibitem[HK86]{HartleKuchar86}
James Hartle and Karel Kuchar.
\newblock Path {I}ntegrals in {P}arametrized {T}heories: {T}he {F}ree
  {R}elativistic {P}article.
\newblock {\em Physical Review D}, 34(2323), 1986.

\bibitem[HT83]{henneauxteitelboim}
Marc Henneaux and Claudio Teitelboim.
\newblock Relativistic quantum mechanics of supersymmetric particles.
\newblock {\em Annals of Physics}, 143, 1983.

\bibitem[JS98]{JoseSaletan:ClassicalMechanics}
Jorge~V. Jos{\'e} and Eugene~J. Saletan.
\newblock {\em Classical {D}ynamics: {A} {C}ontemporary {A}pproach}.
\newblock Cambridge University Press, 1998.

\bibitem[Kle09]{KleinertPathes}
Hagen Kleinert.
\newblock {\em Path Integrals in Quantum Mechanics, Statistics, Polymer
  Physics, and Financial Markets}.
\newblock World Scientific, 2009.

\bibitem[Lan11]{langevin11:twins}
Paul Langevin.
\newblock L'{{\'E}}volution de l'espace et du temps.
\newblock {\em Scientia}, 10, 1911.

\bibitem[Rin06]{Rindler06:Relativity}
Wolfgang Rindler.
\newblock {\em Relativity}.
\newblock Oxford University Press, 2006.

\bibitem[RS93]{Redmount-RQMPath}
Ian~H. Redmount and Wai-Mo Suen.
\newblock Path integration in relativistic quantum mechanics.
\newblock {\em International Journal of Modern Physics A}, 8, 1993.

\bibitem[Tei83]{teitelboim1983}
Claudio Teitelboim.
\newblock Causality {V}ersus {G}auge {I}nvariance in {Q}uantum {G}ravity and
  {S}upergravity.
\newblock {\em Physical Review Letters}, 50, 1983.

\end{thebibliography}

\end{document}